\documentclass[12pt]{article}
\usepackage{psfig}
\usepackage{supertabular}
\newcommand{\sun}{\odot}
\newcommand{\degr}{^\circ}
\begin{document}
\begin{center}
{\LARGE \bf Total masses of the Local Group and M 81 group derived from
			the local Hubble flow}

\bigskip

{\large \bf I.D.Karachentsev$^1$
\ and  O.G.Kashibadze$^2$}

\bigskip
{\large $^1$Special Astrophysical Observatory, Russian Academy of Sciences,
	 N.Arkhyz, KChR, 369167, Russia\\
$^2$ Moscow State University, Moscow, Russia}
\end{center}
\abstract{Based on accurate measurements of distances to nearby galaxies made
with Hubble Space Telescope, we determined the radii of the zero-velocity
surface: $R_0 =0.96\pm0.03$ Mpc for the Local Group and $R_0 =
0.89\pm0.05$ Mpc for the group of galaxies around M~81/M~82. This yields
the total masses of the groups to be $M_T = (1.29\pm0.14) 10^{12} M_{\sun}$
and $M_T = (1.03\pm0.17) 10^{12} M_{\sun}$, respectively.
  The $R_0$--method allowed us to determine the mass ratio of the brightest
two members in the considered groups. Based on the minimum scatter of
galaxies with respect to the Hubble regression, we derived  a mass ratio
of 0.8 : 1.0 for the Milky Way and Andromeda, and 0.54 : 1.00 for
M~82 and M~81, which is quite close to the ratio of luminosities of
these galaxies.
\section{Introduction.}

  Until recently the use of the virial relation  $2T + U = 0$ between
the kinetic $(T)$ and potential $(U)$ energy of a group of galaxies was
the only method of calculating the mass of a system of galaxies on scales
of (0.1 - 1.0) Mpc. However, the uncertainty in membership of some
galaxies even in very nearby groups, the possible absence of a supposed
virial equilibrium and also the unknown character of predominating
motions in groups make virial estimates of the mass not quite reliable
tool, especially in the case of loose groups.

  As has been noted by Lynden-Bell (1981), any group as a large
concentration of mass, has a decelerating effect on the surrounding
Hubble flow. At small distances from the center of a group, the
``velocity - distance'' relation departs from the linear Hubble law
$V = H_0 R$ and crosses the zero-velocity line at a value of $R = R_0$
which was named ``the radius of the zero-velocity surface''.
According to Lynden-Bell, in the case of spherical symmetry the total
mass of the group $M_T$ and the radius $R_0$ are related by a simple
			equation
\begin{equation}
M_T=(\pi^2/8G)\times R_0^3\times T_0^{-2},
\end{equation}
where $G$ is the gravity constant and $T_0$ is the age of the universe.
Thus, the determination of the value of $R_0$ from observational data
enables the mass of the group to be computed since the third parameter
in (1), the age of the universe, is now known with a sufficiently high
accuracy, $T_0 = 13.7\pm0.2$ Gyr (Spergel et al. 2003). These considerations
were used by Sandage (1986, 1987) and Giraud (1990) to estimate the
mass of the Local Group. Based on distances and radial velocities of
a dozen of nearby galaxies, Sandage (1986) calculated the total mass
of the Local Group (LG) to be  $4\times 10^{11} M_{\sun}$. At that time, the
distances to even the nearest galaxies were measured with low accuracy. For
instance, Sandage adopted the following distances: 1.58 Mpc for Leo~A
(best present value 0.69 Mpc), 1.66 Mpc for NGC~300 (best present value
2.15 Mpc), 2.51 Mpc for Pegasus dIr (0.76 Mpc),
2.63 Mpc for NGC~2403 (3.30 Mpc), 5.75 Mpc for M~81
(3.63 Mpc), and 6.31 Mpc for IC~342 (3.28 Mpc). Here we show in brackets
the present-day distances of these galaxies derived from luminosities
of cepheids or the tip of the red giant branch. Recently, Karachentsev
et al. (2002a) have used accurate estimates of distances and radial
velocities for the most nearby 38 galaxies in the neighborhood of the
LG and obtained a value of the total mass of the group to be
$(1.3\pm0.3) 10^{12} M_{\sun}$, i.e. three times as high as the Sandage's
estimate. This approach was applied later by Karachentsev et al.(2002b)
to the determination of $R_0$ and $M_T$ for other nearby groups around
M~81, Centaurus~A, IC~342, NGC~253 (Sculptor filament), and NGC~4736
(Canes Venatici cloud). The summary of the estimates of the total
masses of the groups and their comparison with the virial mass estimates
were presented by Karachentsev (2005). As the comparison showed, in
the groups of galaxies with a "``crosing time'' $T_{cross} < (2 H_0)^{-1}$
the mass estimates from internal (virial) motions and from external
Hubble pattern of velocities are in agreement within the measurement
errors ($\sim$30-40\%). It is evident, however, that the new method deriving
the total mass for a group from $R_0$ needs to be discussed in more detail.

\section{  Observational data and determination of $R_0$.}

Consider a group of galaxies with the center at ``C'' (Figure 1a), which
is located from us (LG) at a distance  $D_c$ and is receding along the
line of sight at a velocity  $V_c$. Let there be a galaxy $G$ in the
vicinity of the group at a distance  $D_g$ from the observer, which is
moving along the line of sight at a velocity  $V_g$. At an angular
separation  $\Theta$ between $C$ and $G$, their mutual linear distance  $R$ is
\begin{equation}
R^2=D^2_g+D^2_c-2D_g\times D_c\times cos\Theta,
\end{equation}
and the mutual velocity difference in projection onto the line between
them is
\begin{equation}
V_{gc}=V_g\times\cos\lambda-V_c\times\cos\mu,
\end{equation}
where $\mu=\lambda+\Theta$,
\begin{equation}
tang\lambda=D_c\times\sin\Theta/(D_g-D_c\times\cos\Theta)
\end{equation}

Here we assumed that random peculiar velocities of the galaxies are low
as compared with the velocities of the regular Hubble flow.

For plotting the relationship between $V_{gc}$ and $R$  we made use of the
data on radial velocities and distances of galaxies in the neighborhood
of the Milky Way, Andromeda (M~31) and M~81. The basic source of data
was the Catalog of Neighboring Galaxies = CNG (Karachentsev et al. 2004)
complemented by the latest measurements of distances to nearby galaxies.
The data used below are collected in Table 1 whose columns contain:
(1,2) galaxy number and name; (3,4) its Galactic coordinates;
(5,6) heliocentric radial velocity and its standard measurement
error; (7) radial velocity reduced to the centroid of the LG
with the apex parameters from NED (NASA Extragalactic Database);
(8,9) measured distance to galaxy and its error (in Mpc);
(10) so-called tidal index or index of galaxy isolation
\begin{equation}
TI_{i}=\max[\log(M_k/D^3_{ik}]+C, \;\;\; k = 1,2,...
\end{equation}
where  $M_k$ is the mass of a neighboring galaxy located at a distance
$D_{ik}$ from the one considered; a constant  $C$  is choosen so that the
negative values of $TI$ correspond to isolated galaxies of the general
field, while the positive values to group members; (11) reference to a
source of data on the galaxy distance. We included into this sample
only the galaxies with accurate measurements of distances, most of the
distances were determined from the tip of the red giant branch (TRGB)
with a typical error of $\sim$10\%. Some preliminary distance estimates made
in the current surveys of nearby galaxies, which are carried out with
ACS at Hubble Space Telescope for the programs of Karachentsev
(\# 9971, \#10235) and Tully (\#10210) are designated as Kar05, Kar06
and Tully05, respectively. Our sample contains a total of 133 nearby
galaxies. About 35 another galaxies are also known in this volume,
but their distances have so far been measured with low accuracy or
else radial velocity measurements are absent.

 Using the velocities and distances from columns (5,6) of Table 1, we
constructed a Hubble diagram  $V_{LG}$ vs. $D_{MW}$ for the nearest 104
galaxies in a sphere of radius 4 Mpc from us. The members of groups
with $TI > 0$ are shown on this diagram (Fig. 2) by open circles, while
the isolated galaxies with $TI < 0$ are presented by filled circles.
The horizontal and vertical bars at them indicate standard errors of
the distance and velocity. The solid line in the figure corresponds to
a Hubble regression with a parameter $H_0 = 72$ km s$^{-1}$ Mpc$^{-1}$ bent at
small distances because of the gravitation deceleration by the mass
of the Local Group. From the family of regressions with the same value
of $H_0$ and different values of $M_{LG}$ (or $R_0$) only one regression is
shown for which the sum of squares of deviations of galaxies is a
minimum. This regression crosses the zero-velocity line at $R_0$ = 0.73 Mpc.

   As can be seen from this diagram, the dispersion of velocities of
galaxies relative to the Hubble regression rises markedly at
$R$ = 3 - 4 Mpc. This is due to galaxies in groups around M~81 and
IC~342/Maffei with high virial motions. Two of them (KDG~63 and KDG~61)
have velocities close to zero with respect to the LG centroid.
Provided that the Hubble diagram is continued to distances $\sim15-20$ Mpc,
we again find there galaxies with velocities  $V_{LG} < 0$, which are
located in the zone of high virial motions in the Virgo cluster.
To elucidate the role played virial motions, we have drawn in Fig. 2
another regression line using the field galaxies only. In the figure
it is displaied by a dotted line, and gives a somewhat smaller value,
$R_0$ = 0.69 Mpc.

\section{The radius  $R_0$ and the LG centroid position.}

As it is known, the Local Group has a binary dumbell-like shape. The
Milky Way with its companions and Andromeda (M 31) with its companions
are separated by a distance of 0.77 Mpc and approaching one another
at a velocity of 123 km s$^{-1}$. We consider below to what extent the
estimate of the zero velocity radius for the LG depends on the position
of its center of mass. We suppose that the center of mass of our
group is on the line connecting the Milky Way and Andromeda at an
arbitrary relative distance  $x = D_c/D_{M31}$. For each value of  $x$  with
a step of 0.05 in the range from 0 (the center resides in the Milky
Way) to 1 ( the center coincides with M~31) the galaxy distances and
velocities were calculated, using expressions (2) and (3). In each case
we constructed a Hubble diagram similar to that shown in Fig.2 and
determined the radius $R_0$ by a $\chi^2$ -- criterion. The results are
presented in Table 2. Column 1 indicates the relative position of the
center of mass, the second and third columns present the value of the
radius  $R_0$ for all galaxies and separately for isolated galaxies.
The fourth column displays the dispersion of velocities relative to
the Hubble regression, and the last one shows the velocity dispersion
after quadratic subtraction of errors, $H_0 \sigma_D$, caused by the
errors of distances. To diminish the contribution of virial motions in
the LG and nearby groups, we computed $R_0$, $\sigma_v$ and $\sigma_{vc}$
in the interval of distances from 0.6 to 2.6 Mpc with respect to the
LG centroid. The data presented in Table 2 allow us to draw the
following conclusions:

  a) The radius of the zero-velocity surface changes in a limited
interval from 0.73 to 0.97 Mpc (0.69 -- 1.02 Mpc for the field galaxies)
at any position of the center of masses of the LG between the Milky Way
and Andromeda, which is evidence of robust estimation of $R_0$ on the
basis of the observational data available.

  b) At different positions of the LG center of masses, the dispersion
of velocities of the galaxies with respect to the Hubble regression
varies from 23 to 61 km s$^{-1}$, and with allowance made for distance
measurement errors $\sigma_{vc}$ varies in the range (13 -- 46) km s$^{-1}$.
For this reason, the Hubble flow around the LG is rather cold.

  c) Adopting a minimum value of $\sigma_v$ or $\sigma_{vc}$ as an indicator
of an optimum position of the LG center of mass, we obtain the value
of  $x = D_c/D_{M31} = 0.55\pm0.05$. This means that the masses of
our Galaxy and M 31 are related as $ M_{MW} : M_{M31} = 0.80 : 1.00$.
This is in good agreement with the ratio of their maximum
rotation velocities $V_m$. According to Fukugita \& Peebles (2004),
$V_m (MW) = 241\pm13$ km s$^{-1}$, while LEDA (Lyon Extragalactic Database)
gives $V_m$ (M~31) = $259\pm5$ km s$^{-1}$. Since the masses (luminosities) of
spiral galaxies are approximately proportional to the cubic power
of $V_m$, then the ratio of masses following from the presented data
is  $M_{MW} : M_{M31} = 0.8 : 1.0$, right the same as our estimate.

  The Hubble diagram for the vicinities of the LG with the position
of the centroid at a distance  $D_c = 0.55\times D_{M31} = 0.42$ Mpc is
presented in Fig. 3. The upper panel of the figure exhibits the
distribution of the galaxies in velocities and distances relative
to the center of masses with indication of errors in measured
velocities and distances. The lower panel contains the original
numbers of all the galaxies in Table 1. As one can see from the
figure, the position of the radius of the zero-velocity surface is
the most sensitive to the velocities and distances of only a few
``strategically situated'' galaxies: Leo~A, WLM, DDO~210, and Sag~DIR.

\section{  $R_0$ and tangential motions}

We have supposed so far tangential velocities of galaxies to be
negligibly small. In order to check the degree this assumption affects
the estimate of the radius $R_0$, we have performed numerical simulations
of the Hubble diagram, adding the tangential component to the radial
velocity of each galaxy from Table 1. The distribution of tangential
velocities was assumed to be Gaussian with a mean  $<V_t>$ and a standard
deviation  $\sigma(V_t)$ = 30 km s$^{-1}$. The orientation of the tangential
velocity vector in positional angle was assigned to be uniformly random.
Assuming different positions of the LG centroid, we performed a lot of
Monte-Carlo simulations at $<V_t>$ = 35 km s$^{-1}$ and $<V_t> =
70 $ km s$^{-1}$. The latter value corresponds to the dispersion of
radial velocities in the LG and M~81 group. The results of determination
of the radius  $R_0$ for 10 series of simulations at $D_c = 0.55 D_{M31}$
are listed in Table 3. As it follows from these data,
the estimates of the radius fluctuate about the mean value
$R_0$ = 0.96 Mpc with a characteristic scatter of 0.03 Mpc.

Returning to the estimate of the LG total mass, we adopt in (1) the age
of the universe to be  $T_0 = 13.7\pm0.2$ Gyr (Spergel et al. 2003), which
leads to an expression
\begin{equation}
M_T/M_{\odot}= 1.46\times 10^{12}\times (R_0/Mpc)^3.
\end{equation}

With the indicated error in the estimate of $T_0$ and the value of the
radius of the zero-velocity surface  $R_0 = (0.96\pm0.03$) Mpc, the
total mass of the Local Group is  $M_T(LG) = (1.29\pm0.14) 10^{12} M_{\sun}$.
Disregarding the mass of other members of the LG, we obtain the values
of the total mass of the two main members of the group: $M_{MW} = 5.8\times 10^{11}M_{\sun}$
and $M_{M31} = 7.1\times 10^{11} M_{\sun}$.

\section{The radius $ R_0$  for M~81 group}

The Hubble diagram for galaxies in the neighborhood of M~81 is presented
in Fig. 4. All the designations here are the same as in Fig. 2. Velocities
and distances are expressed with reference to the group centroid
coincided with the brightest member, M~81. The crossing of the Hubble
regression with the zero-velocity line gives  $R_0 = 1.10$ Mpc. The
regression only for isolated galaxies (dotted curve) results in a
somewhat smaller value, $R_0 = 1.05$ Mpc.

  As it is known, a galaxy, second in luminosity in this group, is M~82
whose radial velocity is by 240 km s${-1}$  higher than in M~81. Considering
that the center of masses of the group lies on the line connecting M~81
and M~82, we determined the velocities and distances of galaxies with
respect to a new center and constructed appropriate Hubble regressions.
The results are tabulated in Table 4. Its first column indicates the
relative position of the center between M~81 $( x = 0)$  and M~82 $( x = 1)$,
the second and the third columns give values of the radius $R_0$ derived
for all galaxies and for the field galaxies only, the last two columns
contain values of the dispersion of velocities with or without taking
account of errors in distances of galaxies, $H_0 \sigma_D$. Estimates of
$R_0$ and $\sigma_v$ were made only over the galaxies situated within
$R$ = 0.6 - 3.0 Mpc from the group center. Two inferences can be made from
the presented data:

  a) The dispersion of velocities of galaxies relative to the Hubble
regression reaches a minimum at the position of the center of masses
at a distance of  $x = 0.35\pm0.05$ from M~81 toward M~82. A mass ratio
of the two galaxies, $M_{M82} : M_{M81} = (0.54\pm0.12)$ : 1.00, corresponds
to this value. The derived mass ratio is consistent, within the errors,
with the ratio of infrared K-luminosities of these galaxies,
$L_{M82} : L_{M81} = 0.47: 1.00$, from the data of the 2MASS survey.

  b) The same as in the case of the LG neighborhood, the galaxies
around the M~81 group demonstrate that the Hubble flow is surprisingly
``cold''. The dispersion of velocities relative to the Hubble regression
at an optimum position ( $x$ = 0.35) of the center of masses of the
group is no higher than 30 km s$^{-1}$, and with allowance made for
measurements errors it drops a few km s$^{-1}$.

  The Hubble diagram for galaxies around M~81 and M~82 at  $x$ = 0.35
is exhibited in Fig. 5. Its upper panel shows errors of the measured
distances and velocities of the galaxies. In the lower panel we indicate
the galaxy numbers labeled by in Table 1. For the determination of the
radius  $R_0$ the most crucial are the positions of the galaxies
UGC~6456 = VII~Zw~403, NGC~4236, KKH~37 and UGC~7242, which makes them
attractive targets for application of more refined methods of distance
estimation.

To define the role of possible tangential motions, we undertook numerical
simulations of the Hubble flow around M~81/M~82 given the same parameters
as in the case of the LG. At a mean tangential velocity of the galaxies
$<V_t>$ = 35 km s$^{-1}$, we obtained actually the same position of the center
of mass, $< x > = 0.35\pm0.03$ and the mean value of the radius  $< R_0> =
0.89\pm0.05$ Mpc. The simulations made with $< V_t> = 70$ km s$^{-1}$ left these
parameters almost unchanged but increased their errors: $< x > =
0.37\pm0.05$ and $< R_0> = 0.87\pm0.10$ Mpc. Adopting for $R_0$ the value of
$0.89\pm0.05$ Mpc, we obtain an estimate of the total mass of the group
$M_T = (1.03\pm0.17) 10^{12} M_{\sun}$. Then, neglecting the contribution of
other members of the group, we derive individual masses $M_{M81} =
6.7\times 10^{11} M_{\sun}$ and $M_{M82} = 3.6\times 10^{11} M_{\sun}$ for the brightest two
galaxies.

\section{ Another application of the  $R_0$-- method.}

P.J.E.Peebles (2005) directed our attention to the existence of
another approach to estimating the radius of the zero-velocity surface
for a group or cluster. Let a galaxy  $G$ falls radially on the center of
a group  $C$  at a velocity  $V_i$ ( see Fig.1b). If the center of the
group is moving away from us at a velocity $V_c$ directed along the line
of sight, then the velocity along the line of sight of the galaxy  $G$
which is located at an angular separation  $\Theta$ from the group center
will be

 \begin{equation}
V_g=V_c\times\cos\Theta - V_i\times \cos\lambda.
\end{equation}

The falling velocity of the galaxy is then expressed as
 \begin{equation}
    V_i=[V_c\times\cos\Theta- V_g]/\cos\lambda,
\end{equation}
and it is precisely this velocity should be compared with the distance
of the galaxy from the group center found from (2) when describing the
pattern of motions of galaxies around a massive group. When the angles
$\lambda$ and $\Theta$ are small (i.e. the galaxy is located strictly in front
or behind the group center), equations (3) and (8) yield an about the
same infall velocity toward the group center. Apparently, at
angles $\lambda$ close to 90$\degr$ the discrepancy between the
two approaches
becomes significant.

  In order to determine the radius  $R_0$ by the new method, we plotted
a Hubble diagram for the Local Group and its neighborhood at the
position of the center of masses on  $x = D_c/D_{M31}$ =0.55. Using
expression (8) and excluding galaxies with   $\cos\lambda < 0.7$,
we obtained a value  $R_0 = 0.92$ Mpc (and 0.78 Mpc for field galaxies
with $TI < 0)$. The dispersion of velocities relative to the Hubble
regression proven to be equal to 29 km s$^{-1}$, or 18 km s$^{-1}$ after taking
account of errors in galaxy distances. As one can see, in the case of
the LG the differences in the estimates of  $R_0$ and $\sigma_v$ for the
two approaches turn out to be small. However, exploring the Hubble
flow around the M~81 group by this method, we found considerable
discrepancies. At the former of the center of masses,  $x = 0.35$, and
with the exclusion of galaxies having  $\cos\lambda <0.7$, we obtained
a significantly larger radius  $R_0 = 1.31$ Mpc (or 1.23 Mpc for the
field galaxies). The scatter of galaxies on the Hubble diagram also
increased, making  $\sigma_v = 70$ km s$^{-1}$ and  $\sigma_{vc}$ = 35 km s$^{-1}$

  A direct comparison of the two discussed methods of estimating the
radius $R_0$ is now impeded because of absence of observational data on
tangential velocities of galaxies. However, such data may be available
in the near future after completion of cosmic projects like SIM
(Space Interferometric Mission) described by Peebles et al. (2001).
In the absence of data on space vectors of galaxy velocities these
two methods lay actually different emphasis on the properties of the
Hubble flow in the vicinity of nearby groups. In the former case we
supposed that most of the galaxies under study are not in the ``infall
zone'' but on the asimptotic Hubble relationship (the model of the
minor attractor). The latter approach assumes that numerous galaxies
being discussed are involved in the infall zone (the model of
the major extended attractor). The Hubble diagrams in Fig.3 and 5
sugest the former approach to be preferred.

\section{ Concluding remarks }

The measurements of distances of many nearby galaxies accomplished
during the last 2--3 years to an accuracy of $\sim$10\% served us an
observational basis for determination of the masses of nearby groups
not from internal (virial) motions, but from the external Hubble
field of velocities around the groups. The application of this method
put forward by Lynden-Bell (1981) and Sandage (1986) assume the
following conditions to be satisfied: 1) spherically symmetric shape
of the group potential well, 2) small random motions of galaxies
relative the regular Hubble flow, and 3) sufficiently high number
density of test particles (galaxies) for which radial velocities and
distances are known with high accuracy.

  Based on the most recent measurements of distances to galaxies made
with Hubble Space Telescope, we determined the radii of the zero-velocity
surface: $R_0 =0.96\pm0.03$ Mpc for the LG and  $R_0 = 0.89\pm0.05$ Mpc for
the group of galaxies around M~81/M~82. With the errors indicated, the
formal accuracy of estimation of the total mass of the groups by the
new method is only $\sim$15\%, which is by about a factor three better than
from virial motions. However, the  $R_0$--method is probably to contain
systematic errors which need a special study.

  At integrated luminosities  $L_B = 10.1\times 10^{10} L_{\sun}$ for the LG and
$L_B = 6.1\times 10^{10} L_{\sun}$ for the M~81 group (Karachentsev, 2005), their
total mass-to-blue luminosity ratios make only (12.8$\pm1.4) M_{\sun}/L_{\sun}$
and $(16.9\pm2.8) M_{\sun}/L_{\sun}$, respectively. The obtained values of
$M_T/L_B$ are much lower than the old virial estimates $M_T/L_B  \sim
100 M_{\sun}/L_{\sun}$ (Tully, 1987), which were considered to be typical of
poor groups of galaxies. Since more than half of galaxies in the Local
volume are members of such groups, this results in a rather low average
density of matter in the Local volume.

  Random motions of galaxies relative to the regular Hubble flow makes
only 15 - 25 km s$^{-1}$ within $(1 - 3) R_0$ around the LG and  3--  28 km s$^{-1}$
in a similar zone around M~81. The observed ``coldness'' of the local
Hubble flow is independent evidence of low density of the part of
matter in the Local volume which concentrates in groups.

  It is also interest that the $R_0$--method made it possible to
determine the mass ratio in the brightest two members of the discussed
groups. Based on the minimum scatter of galaxies with respect to the
Hubble regression, we derived a mass ratio of 0.8 : 1.0 for the Milky
Way and Andromeda, and found a ratio of masses of  0.54 : 1.00 for
M~82 and M~81, which is quite close to the ratio of luminosities of
these galaxies.

It is pleasure to thank P.J.E. Peebles for very useful discussions.
We used in this work the NASA Extragalactic Database (NED), Lyon
Extragalactic Database (LEDA) and the data of the 2 Micron All Sky
Survey (2MASS). The work was supported through grant of RFBR 04--02--16115
and grant DFG--RFBR 02--02--04012.

{}

\onecolumn
\tablecaption{Galaxies with accurate distances and radial velocities
	   in/around the Local Group and the M81 group.}
\tablehead{\hline
\multicolumn{1}{c}{N} &
\multicolumn{1}{c}{Name}   &
\multicolumn{1}{c}{$l$}  &
\multicolumn{1}{c}{$b$}&
\multicolumn{1}{c}{$V_h$}&
\multicolumn{1}{c}{$\pm dV$}&
\multicolumn{1}{c}{$V_{LG}$} &
\multicolumn{1}{c}{$D_{MW}$}&
\multicolumn{1}{c}{$\pm dD$}  &
\multicolumn{1}{c}{$TI$} &
\multicolumn{1}{c}{Reference}  \\
 & &\multicolumn{1}{c}{deg} &
\multicolumn{1}{c}{deg}  &
\multicolumn{2}{c}{km/s} &
\multicolumn{1}{c}{km/s}  &
\multicolumn{2}{c}{Mpc}   &   &                 \\
\hline}
\par
\begin{supertabular}{rlrrrrrcrrl}
  1 & WLM      &  75.86 &$-$73.62 & $-$116 & 2 &  $-$10 &   0.92& 0.04 &  0.3 & CNG   \\
  2 & ESO349-31& 351.48 &$-$78.12 &  207   & 7 &  216 &   3.21  & 0.31 &  0.5 & Kar05  \\
  3 & NGC55    & 332.67 &$-$75.74 &  129   & 3 &  111 &   2.12  & 0.21 & $-$0.4 & Seth et al.05 \\
  4 & IC10     & 118.97 &$-$3.34 & $-$344  & 1 &  $-$60 &   0.66& 0.06 &  1.8 & CNG \\
  5 & ESO294-10& 320.42 &$-$74.42 &  117   & 5 &   81 &   1.92  & 0.19 &  1.0 & CNG \\
  6 & NGC147   & 119.82 &$-$14.25 & $-$193 & 3 &   85 &   0.76  & 0.08 &  3.0 & CNG \\
  7 & AndIII   & 119.37 &$-$26.26 & $-$355 & 9 &  $-$92 &   0.76& 0.07 &  3.5 & CNG \\
  8 & NGC185   & 120.79 &$-$14.48 & $-$202 & 3 &   73 &   0.62  & 0.06 &  2.3 & CNG \\
  9 & NGC205   & 120.72 &$-$21.14 & $-$244 & 3 &   24 &   0.83  & 0.11 &  3.7 & CNG \\
 10 & NGC221   & 121.15 &$-$21.98 & $-$145 & 6 &  121 &   0.77  & 0.04 &  6.8 & CNG \\
 11 & M31      & 121.17 &$-$21.57 & $-$301 & 1 &  $-$35 &   0.77& 0.04 &  4.6 & CNG \\
 12 & AndI     & 121.68 &$-$24.82 & $-$380 &11 & $-$120 &   0.81& 0.03 &  3.7 & CNG \\
 13 & NGC247   & 113.94 &$-$83.56 &  160   & 2 &  215 &   3.65  & 0.38 &  1.3 & Kar05 \\
 14 & NGC253   &  97.43 &$-$87.97 &  241   & 2 &  274 &   3.94  & 0.37 &  0.3 & CNG \\
 15 & DDO6     & 119.39 &$-$83.88 &  295   & 5 &  348 &   3.34  & 0.24 &  0.5 & CNG \\
 16 & SMC      & 302.81 &$-$44.33 &  158   & 4 &  $-$22 &   0.06& 0.01 &  3.5 & CNG \\
 17 & NGC300   & 299.21 &$-$79.42 &  144   & 5 &  114 &   2.15  & 0.10 & $-$0.3 & CNG \\
 18 & Sculptor & 287.53 &$-$83.16 &  110   & 1 &   96 &   0.09  & 0.01 &  2.8 & CNG \\
 19 & LGS-3    & 126.77 &$-$40.88 & $-$286 & 5 &  $-$74 &   0.62& 0.02 &  1.7 & CNG \\
 20 & IC1613   & 129.79 &$-$60.56 & $-$232 & 1 &  $-$89 &   0.73& 0.02 &  0.9 & CNG \\
 21 & KKH5     & 125.49 &$-$11.35 &   39   & 2 &  304 &   4.26  & 0.43 & $-$1.2 & CNG \\
 22 & NGC404   & 127.03 &$-$27.01 & $-$48  & 9 &  195 &   3.06  & 0.37 & $-$1.0 & CNG \\
 23 & AndV     & 126.22 &$-$15.12 & $-$403 & 4 & $-$143 &   0.81& 0.04 &  2.8 & CNG \\
 24 & AndII    & 128.92 &$-$29.16 & $-$188 & 3 &   46 &   0.68  & 0.02 &  2.4 & CNG \\
 25 & M33      & 133.61 &$-$31.33 & $-$180 & 3 &   36 &   0.85  & 0.04 &  2.0 & CNG \\
 26 & KKH6     & 129.68 &$-$10.21 &   17   & 1 &  270 &   3.73  & 0.38 & $-$0.8 & Kar05 \\
 27 & NGC625   & 273.67 &$-$73.12 &  405   & 1 &  335 &   3.89  & 0.39 & $-$0.4 & Cannon et al.05 \\
 28 & Phoenix  & 272.16 &$-$68.95 & $-$13  &29 & $-$106 &   0.44& 0.02 &  0.8 & CNG \\
 29 & Maffei1  & 135.86 &$-$0.55 &   66    &22 &  297 &   3.01  & 0.60 &  2.7 & Fingerhut et.04  \\
 30 & Fornax   & 237.29 &$-$65.65 &   53   & 9 &  $-$32 &   0.14& 0.01 &  2.3 & CNG \\
 31 & KK35     & 138.20 & 10.30 &  105     & 1 &  320 &   3.16  & 0.32 &  2.4 & CNG \\
 32 & IC342    & 138.17 & 10.58 &   31     & 3 &  245 &   3.28  & 0.27 & $-$0.1 & CNG     \\
 33 & UGCA86   & 139.77 & 10.64 &   67     & 4 &  275 &   2.96  & 0.31 &  0.3 & Kar05   \\
 34 & CamA     & 137.25 & 16.20 &  $-$47   & 1 &  164 &   3.93  & 0.47 &  0.1 & CNG     \\
 35 & UGCA92   & 144.70 & 10.51 &  $-$99   & 5 &   89 &   3.01  & 0.31 &  1.1 & Kar05   \\
 36 & NGC1560  & 138.37 & 16.02 &  $-$36   & 5 &  171 &   3.45  & 0.36 &  1.0 & CNG     \\
 37 & CamB     & 143.38 & 14.42 &    77    & 5 &  266 &   3.34  & 0.32 &  1.0 & CNG     \\
 38 & UGCA105  & 148.52 & 13.66 &  111     & 5 &  279 &   3.15  & 0.32 &  0.3 & CNG     \\
 39 & LMC      & 280.47 &$-$32.89 &   278  & 2 &   28 &   0.05  & 0.01 &  3.6 & CNG     \\
 40 & KKH34    & 140.42 & 22.35 &  110     & 1 &  299 &   4.61  & 0.46 & $-$0.8 & CNG     \\
 41 & Carina   & 260.11 &$-$22.22 &   223  &60 &  $-$53 & 0.10  & 0.01 &  2.7 & CNG     \\
 42 & KKH37    & 133.98 & 26.54 &    10    & 1 &  214 &   3.39  & 0.33 & $-$0.3 & Kar05     \\
 43 & HIZSS003 & 217.71 &   0.09 &   280   & 1 &  101 &   1.69  & 0.17 & $-$0.6 & Silva et al.05\\
 44 & NGC2366  & 146.43 & 28.53 &   99     & 3 &  253 &   3.19  & 0.41 &  1.0 & CNG \\
 45 & NGC2403  & 150.57 & 29.19 &  131     & 3 &  268 &   3.30  & 0.36 &  0.0 & CNG \\
 46 & HoII     & 144.28 & 32.69 &  157     & 1 &  311 &   3.39  & 0.20 &  0.6 & CNG \\
 47 & KDG52    & 143.82 & 33.01 &  113     & 5 &  268 &   3.55  & 0.26 &  0.7 & CNG \\
 48 & DDO53    & 149.30 & 34.95 &    20    & 1 &  151 &   3.56  & 0.24 &  0.7 & CNG \\
 49 & UGC4483  & 144.97 & 34.38 &  156     & 1 &  304 &   3.21  & 0.18 &  0.5 & CNG \\
 50 & HoI      & 140.73 & 38.65 &  139     & 1 &  291 &   3.84  & 0.46 &  1.5 & CNG \\
 51 & NGC2976  & 143.92 & 40.90 &     3    & 5 &  139 &   3.56  & 0.38 &  2.7 & CNG \\
 52 & BK3n     & 142.31 & 40.83 &  $-$40   & 5 &  101 &   4.02  & 0.26 &  1.0 & CNG \\
 53 & M81      & 142.09 & 40.90 &  $-$35   & 4 &  107 &   3.63  & 0.34 &  2.2 & CNG \\
 54 & M82      & 141.40 & 40.57 &  202     & 4 &  347 &   3.53  & 0.26 &  2.7 & CNG \\
 55 & KDG61    & 142.50 & 41.28 & $-$116   &30 &   23 &   3.60  & 0.25 &  3.9 & CNG \\
 56 & A0952+69 & 141.74 & 40.92 &  100     & 5 &  243 &   3.87  & 0.21 &  1.9 & CNG \\
 57 & LeoA     & 196.90 & 52.42 &    24    & 4 &  $-$40 &   0.69& 0.06 &  0.2 & CNG \\
 58 & SexB     & 233.20 & 43.78 &  301     & 1 &  111 &   1.36  & 0.07 & $-$0.7 & CNG \\
 59 & NGC3109  & 262.10 & 23.07 &  403     & 1 &  110 &   1.33  & 0.08 & $-$0.1 & CNG \\
 60 & NGC3077  & 141.90 & 41.66 &    13    & 4 &  153 &   3.82  & 0.38 &  1.9 & CNG \\
 61 & Antlia   & 263.10 & 22.31 &  362     & 1 &   66 &   1.32  & 0.06 &  2.3 & CNG \\
 62 & KDG63    & 144.13 & 43.10 & $-$129   & 5 &    0 &   3.50  & 0.24 &  1.8 & CNG \\
 63 & LeoI     & 225.98 & 49.11 &  285     & 2 &  128 &   0.25  & 0.02 &  1.5 & CNG \\
 64 & SexA     & 246.15 & 39.88 &  324     & 1 &   94 &   1.32  & 0.04 & $-$0.6 & CNG \\
 65 & SexdSph  & 243.50 & 42.27 &  226     & 2 &    8 &   0.09  & 0.01 &  2.8 & CNG \\
 66 & HS117    & 138.14 & 41.30 &  $-$37   & 5 &  116 &   3.96  & 0.39 &  1.9 & Kar05\\
 67 & DDO78    & 141.14 & 44.00 &    55    & 9 &  191 &   3.72  & 0.26 &  1.8 & CNG \\
 68 & IC2574   & 140.20 & 43.60 &    57    & 2 &  197 &   4.02  & 0.41 &  0.9 & CNG \\
 69 & DDO82    & 137.90 & 42.18 &    56    & 3 &  207 &   4.00  & 0.40 &  0.9 & CNG \\
 70 & KDG73    & 136.88 & 44.23 &  116     & 6 &  263 &   3.70  & 0.22 &  1.3 & CNG \\
 71 & LeoII    & 220.16 & 67.23 &    76    & 5 &  $-$18 &   0.21& 0.02 &  1.7 & CNG \\
 72 & UGC6456  & 127.84 & 37.33 & $-$103   & 1 &   89 &   4.34  & 0.43 & $-$0.3 & CNG \\
 73 & UGC6541  & 151.90 & 63.27 &  250     & 4 &  304 &   3.89  & 0.47 & $-$0.7 & CNG \\
 74 & NGC3738  & 144.56 & 59.32 &  228     & 4 &  305 &   4.90  & 0.49 & $-$1.0 & CNG \\
 75 & NGC3741  & 157.57 & 66.45 &  230     & 4 &  264 &   3.03  & 0.30 & $-$0.8 & CNG \\
 76 & KK109    & 156.85 & 68.98 &  212     & 1 &  241 &   4.51  & 0.45 & $-$0.6 & CNG \\
 77 & DDO99    & 166.20 & 72.75 &  242     & 1 &  248 &   2.64  & 0.21 & $-$0.5 & CNG \\
 78 & NGC4068  & 138.91 & 63.04 &  210     & 3 &  290 &   4.31  & 0.42 & $-$1.0 & Kar05 \\
 79 & NGC4163  & 163.21 & 77.70 &  163     & 5 &  164 &   2.96  & 0.29 &  0.1 & Kar05 \\
 80 & ESO321-14& 294.85 & 24.05 &  613     & 5 &  337 &   3.19  & 0.26 & $-$0.3 & CNG    \\
 81 & UGC7242  & 128.87 & 50.60 &   68     & 2 &  213 &   5.42  & 0.52 &  0.4 & Kar05  \\
 82 & KDG90    & 161.10 & 78.06 &  280     & 6 &  283 &   2.86  & 0.14 &  1.6 & CNG    \\
 83 & NGC4214  & 160.26 & 78.07 &  291     & 3 &  295 &   2.94  & 0.18 & $-$0.7 & CNG    \\
 84 & UGC7298  & 135.22 & 64.06 &  173     & 1 &  255 &   4.21  & 0.42 & $-$0.7 & CNG    \\
 85 & NGC4236  & 127.41 & 47.36 &     0    & 4 &  160 &   4.45  & 0.44 & $-$0.4 & CNG    \\
 86 & NGC4244  & 154.56 & 77.16 &  243     & 1 &  255 &   4.49  & 0.45 &  0.0 & CNG    \\
 87 & IC3104   & 301.41 &$-$16.95 &   430  & 5 &  171 &   2.27  & 0.19 & $-$0.5 & CNG    \\
 88 & NGC4395  & 162.08 & 81.54 &  320     & 1 &  315 &   4.61  & 0.46 &  0.1 & CNG    \\
 89 & DDO126   & 148.60 & 78.74 &  218     & 5 &  231 &   4.87  & 0.49 &  0.1 & CNG    \\
 90 & DDO125   & 137.75 & 72.94 &  195     & 4 &  240 &   2.54  & 0.17 & $-$0.9 & CNG    \\
 91 & NGC4449  & 136.85 & 72.40 &  201     & 4 &  249 &   4.21  & 0.42 &  0.0 & CNG    \\
 92 & UGC7605  & 150.99 & 80.13 &  310     & 1 &  317 &   4.43  & 0.44 &  0.7 & CNG    \\
 93 & NGC4605  & 125.33 & 55.47 &  143     & 5 &  276 &   5.47  & 0.53 & $-$1.1 & Kar05  \\
 94 & IC3687   & 131.95 & 78.46 &  358     & 1 &  385 &   4.57  & 0.46 &  1.1 & CNG    \\
 95 & NGC4736  & 123.36 & 76.01 &  309     & 1 &  353 &   4.66  & 0.47 & $-$0.5 & CNG    \\
 96 & GR8      & 310.74 & 76.98 &  214     & 3 &  136 &   2.10  & 0.34 & $-$1.2 & CNG    \\
 97 & IC4182   & 107.71 & 79.09 &  320     & 1 &  356 &   4.70  & 0.65 &  0.6 & CNG    \\
 98 & DDO165   & 120.75 & 49.36 &    31    & 1 &  196 &   4.57  & 0.46 &  0.0 & CNG    \\
 99 & UGC8215  & 114.58 & 70.03 &  218     & 1 &  297 &   4.55  & 0.45 & $-$0.5 & Kar05  \\
100 & DDO167   & 111.62 & 70.32 &  163     & 6 &  243 &   4.19  & 0.47 &  0.0 & CNG    \\
101 & DDO168   & 110.76 & 70.66 &  194     & 1 &  273 &   4.33  & 0.43 &  0.0 & CNG    \\
102 & NGC5102  & 309.73 & 25.84 &  467     & 7 &  230 &   3.40  & 0.39 &  0.7 & CNG    \\
103 & NGC5204  & 113.50 & 58.01 &  203     & 1 &  341 &   4.65  & 0.46 & $-$1.1 & CNG    \\
104 & UGC8508  & 111.14 & 61.31 &   62     & 5 &  186 &   2.56  & 0.15 & $-$1.0 & CNG    \\
105 & NGC5237  & 311.88 & 19.22 &  361     & 4 &  122 &   3.33  & 0.33 &  2.1 & Kar05  \\
106 & UGC8638  &  23.28 & 78.99 &  274     & 1 &  273 &   4.27  & 0.40 & $-$1.3 & Kar05  \\
107 & DDO181   &  89.73 & 73.12 &  202     & 1 &  272 &   3.02  & 0.31 & $-$1.3 & Tully05\\
108 & ESO325-11& 313.51 & 19.91 &  540     & 4 &  307 &   3.40  & 0.39 &  1.1 & CNG    \\
109 & DDO183   &  77.79 & 73.45 &  191     & 1 &  257 &   3.18  & 0.32 & $-$0.8 & Tully05\\
110 & KKH86    & 339.04 & 62.60 &  287     & 3 &  209 &   2.61  & 0.16 & $-$1.5 & CNG    \\
111 & UGC8833  &  69.71 & 73.96 &  226     & 5 &  285 &   3.12  & 0.31 & $-$1.4 & Tully05\\
112 & KK230    &  63.71 & 71.99 &    62    & 2 &  126 &   1.92  & 0.18 & $-$1.0 & Kar05  \\
113 & DDO187   &  25.57 & 70.46 &  152     & 4 &  172 &   2.28  & 0.22 & $-$1.3 & Tully05\\
114 & DDO190   &  82.01 & 64.48 &  150     & 4 &  263 &   2.79  & 0.26 & $-$1.3 & CNG    \\
115 & UMin     & 104.95 & 44.80 & $-$247   & 1 &  $-$44 &   0.06& 0.01 &  3.3 & CNG    \\
116 & ESO274-01& 326.80 &  9.33 &  522     & 5 &  335 &   3.12  & 0.30 & $-$1.0 & Kar06  \\
117 & KKR25    &  83.88 & 44.41 & $-$139   & 2 &   68 &   1.86  & 0.12 & $-$0.7 & CNG    \\
118 & Draco    &  86.36 & 34.75 & $-$293   &21 &  $-$48 &   0.08& 0.01 &  3.0 & CNG    \\
119 & MilkyWay &   0.73 &  0.57 &     0    &10 &  $-$88 &   0.01& 0.00 &  2.5 & CNG    \\
120 & IC4662   & 328.55 &$-$17.85 &   308  & 4 &  145 &   2.44  & 0.24 & $-$0.9 & Kar05  \\
121 & NGC6503  & 100.57 & 30.64 &    43    & 7 &  301 &   5.27  & 0.53 & $-$1.2 & CNG    \\
122 & SagdSph  &   5.61 &$-$14.09 &   142  & 4 &  161 &   0.02  & 0.00 &  5.6 & CNG    \\
123 & NGC6789  &  94.97 & 21.52 & $-$141   & 9 &  144 &   3.60  & 0.36 & $-$1.4 & CNG    \\
124 & SagDIG   &  21.06 &$-$16.28 &   $-$77& 4 &   23 &   1.04  & 0.07 & $-$0.3 & CNG    \\
125 & NGC6822  &  25.34 &$-$18.40 & $-$57  & 2 &   64 &   0.50  & 0.01 &  0.6 & CNG    \\
126 & DDO210   &  34.05 &$-$31.34 &  $-$137& 5 &   13 &   0.94  & 0.04 & $-$0.1 & CNG    \\
127 & IC5152   & 343.92 &$-$50.19 &  124   & 3 &   75 &   2.07  & 0.18 & $-$1.1 & CNG    \\
128 & UGCA438  &  11.86 &$-$70.86 &    62  & 5 &   99 &   2.23  & 0.15 & $-$0.7 & CNG    \\
129 & CasdSph  & 109.46 &$-$9.96 & $-$307  & 2 &   $-$5 &   0.79& 0.04 &  2.0 & CNG    \\
130 & Pegasus  &  94.77 &$-$43.55 & $-$184 & 2 &   60 &   0.76  & 0.08 &  1.2 & CNG    \\
131 & KKH98    & 109.09 &$-$22.38 & $-$137 & 3 &  151 &   2.45  & 0.13 & $-$0.7 & CNG    \\
132 & PegdSph  & 106.04 &$-$36.32 & $-$354 & 3 &  $-$94 &   0.82& 0.02 &  1.7 & CNG    \\
133 & NGC7793  &   4.52 &$-$77.17 &  229   & 2 &  252 &   3.91  & 0.41 &  0.1 & CNG    \\
\hline
\par
\end{supertabular}

\twocolumn
\begin{table}
\caption{The Hubble flow parameters for the Local Group.}
\vspace{0.5cm}
\begin{tabular}{lcccc} \hline
Centroid   &    $ R_0$  &  $R_0$,field &     $\sigma_v$ & $\sigma_{vc}$ \\
position   &          &            &             &           \\
\hline
	   &      Mpc &   Mpc      &       km/s  &  km/s     \\
\hline
0.00       &     0.73 &   0.69     &       31.9  &  20.5     \\
0.05       &     0.76 &   0.70     &       31.4  &  19.8     \\
0.10       &     0.77 &   0.71     &       30.6  &  19.0     \\
0.15       &     0.78 &   0.73     &       29.9  &  18.3     \\
0.20       &     0.79 &   0.74     &       29.4  &  17.8     \\
0.25       &     0.79 &   0.75     &       29.1  &  17.2     \\
0.30       &     0.81 &   0.77     &       28.9  &  17.0     \\
0.35       &     0.83 &   0.79     &       28.6  &  16.8     \\
0.40       &     0.83 &   0.80     &       28.3  &  16.7     \\
0.45       &     0.85 &   0.81     &       28.1  &  16.6     \\
0.50       &     0.86 &   0.83     &       27.8  &  16.3     \\
0.55       &     0.95 &   0.89     &       23.8  &  12.8     \\
0.60       &     0.90 &   0.89     &       37.7  &  22.4     \\
0.65       &     0.92 &   0.91     &       38.5  &  23.2     \\
0.70       &     0.93 &   0.92     &       39.6  &  24.3     \\
0.75       &     0.94 &   0.94     &       40.5  &  25.2     \\
0.80       &     0.96 &   0.96     &       41.7  &  26.4    \\
0.85       &     0.97 &   0.97     &       42.8  &  27.5     \\
0.90       &     0.97 &   0.99     &       50.3  &  34.3      \\
0.95       &     0.81 &   1.01     &       59.7  &  44.0     \\
1.00       &     0.88 &   1.02     &       61.0  &  46.3     \\
\hline
\end{tabular}
\end{table}

\begin{table}
\caption{The radius $R_0$ for the Local Group
	   for x = 0.55 and tangential motions.}
\vspace{0.5cm}
\begin{tabular}{cc} \hline
      $  R_0$ (Mpc)     &    $ R_0$ (Mpc)     \\
       $ (V_t$ = 35 km/s)&    $(V_t$ = 70 km/s) \\
\hline
	0.99      &          0.89               \\
	0.96      &          1.03               \\
	0.95      &          0.96               \\
	0.96      &          0.98               \\
	0.95      &          0.91               \\
	0.94      &          0.95               \\
	0.99      &          1.02               \\
	0.96      &          0.97               \\
	0.91      &          0.98               \\
	0.95      &          0.95               \\
\hline
Mean  0.96$\pm$0.02 &        0.96$\pm$0.04      \\
\hline
\end{tabular}
\end{table}

\begin{table}
\caption{The Hubble flow parameters around M81 group.}
\vspace{0.5cm}
\begin{tabular}{lcccr} \hline

Centroid &  $R_0$   &$R_0$,field & $\sigma_v$ & $\sigma_{vc}$   \\
position &  (Mpc) & (Mpc)    & (km/s)  &  (km/s)     \\
\hline
0.00     &  1.10  &  1.05    &  49.9   & 22.2        \\
0.05     &  1.07  &  1.03    &  45.4   & 17.7        \\
0.10     &  1.04  &  1.01    &  41.2   & 13.5        \\
0.15     &  1.01  &  0.98    &  37.2   &  9.7        \\
0.20     &  0.98  &  0.96    &  33.7   &  6.2        \\
0.25     &  0.95  &  0.94    &  31.0   &  3.4        \\
0.30     &  0.92  &  0.92    &  29.1   &  1.5        \\
0.35     &  0.89  &  0.90    &  28.2   &  0.6        \\
0.40     &  0.83  &  0.86    &  30.1   &  2.5        \\
0.50     &  0.80  &  0.84    &  32.6   &  5.0        \\
0.55     &  0.79  &  0.82    &  35.7   &  8.2        \\
0.60     &  0.77  &  0.80    &  39.4   & 11.8        \\
0.65     &  0.74  &  0.78    &  43.6   & 16.1        \\
0.70     &  0.71  &  0.76    &  48.2   & 20.6        \\
0.75     &  0.69  &  0.74    &  53.0   & 25.5        \\
0.80     &  0.66  &  0.72    &  58.0   & 30.5        \\
0.85     &  0.63  &  0.70    &  63.0   & 35.5        \\
0.90     &  0.60  &  0.69    &  68.4   & 40.8        \\
0.95     &  0.63  &  0.67    &  73.8   & 46.3        \\
1.00     &  0.60  &  0.65    &  79.4   & 51.9        \\
\hline
\end{tabular}
\end{table}

\onecolumn

  \begin{figure}[hbt]
 \vbox{\includegraphics{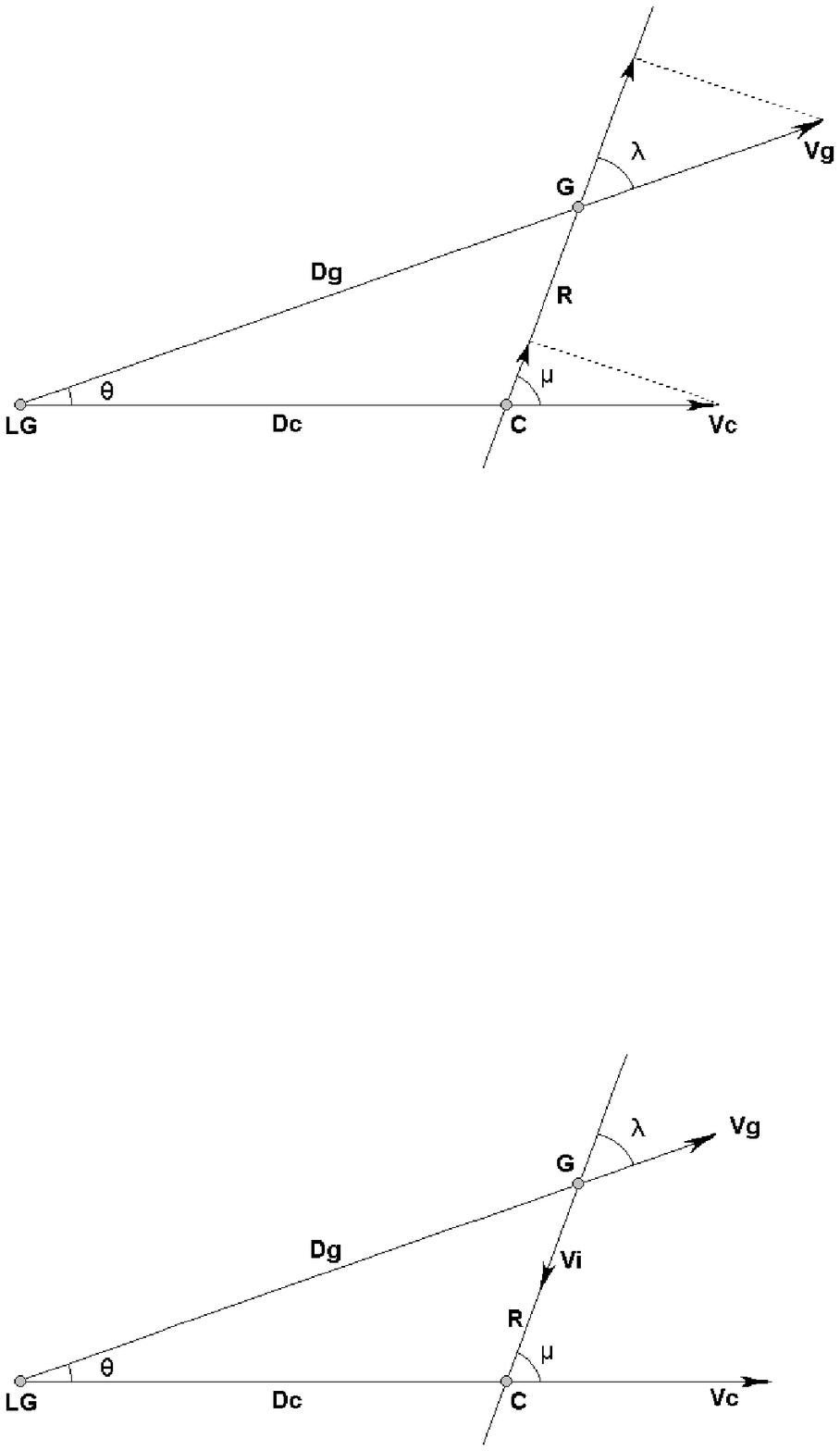}}
\vspace{20cm}
\caption{Diagrams of motion of a galaxy relative to group center.}
\end{figure}

\begin{figure}
 \vbox{\includegraphics{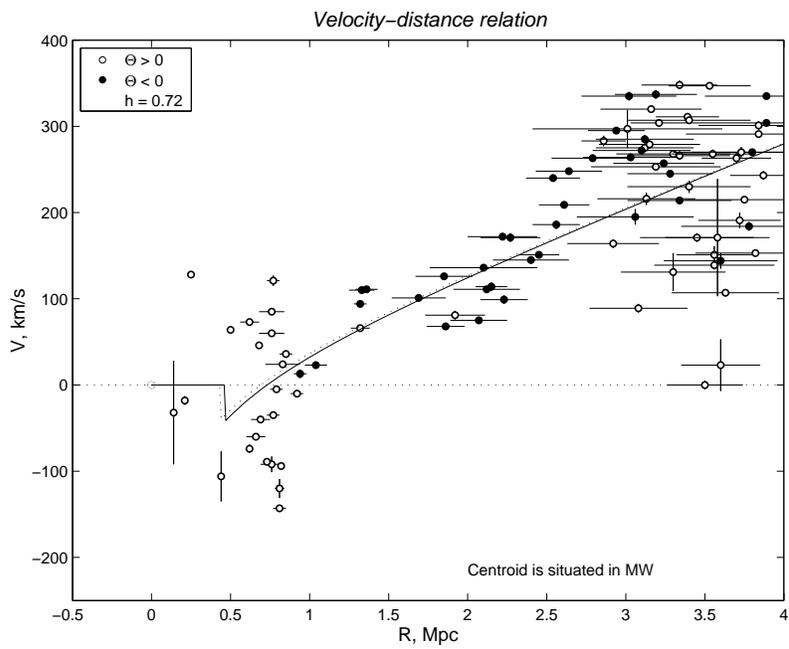}}
\vspace{15cm}
\caption{Hubble diagram for the Local Group with the center in the
	Milky Way.}
\vspace{15cm}
\end{figure}

  \begin{figure}[hbt]
 \vbox{\includegraphics{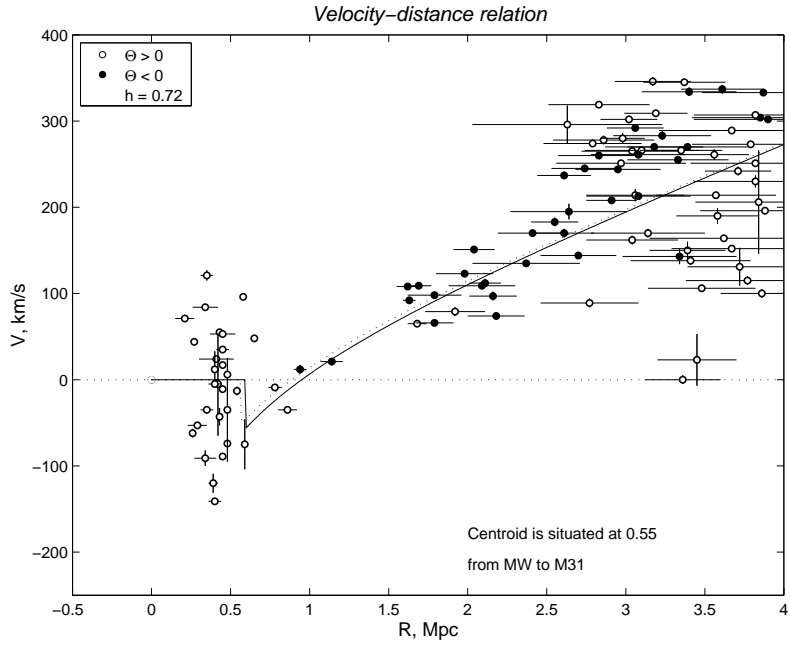}}
 \vbox{\includegraphics{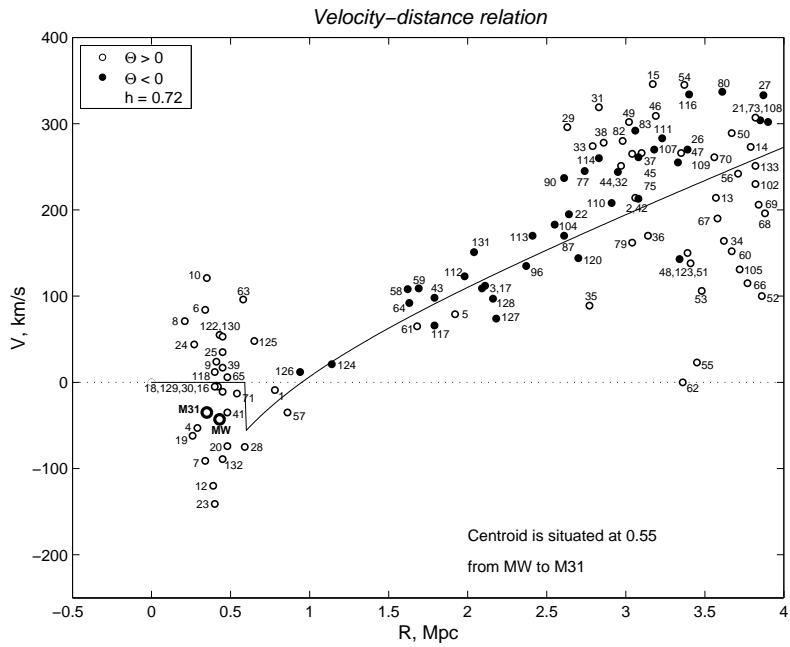}}
\vspace{20cm}
\caption{Hubble diagram for the Local Group with the center of
masses at $x=0.55$ towards M~31.}
\end{figure}

  \begin{figure}[hbt]
 \vbox{\includegraphics{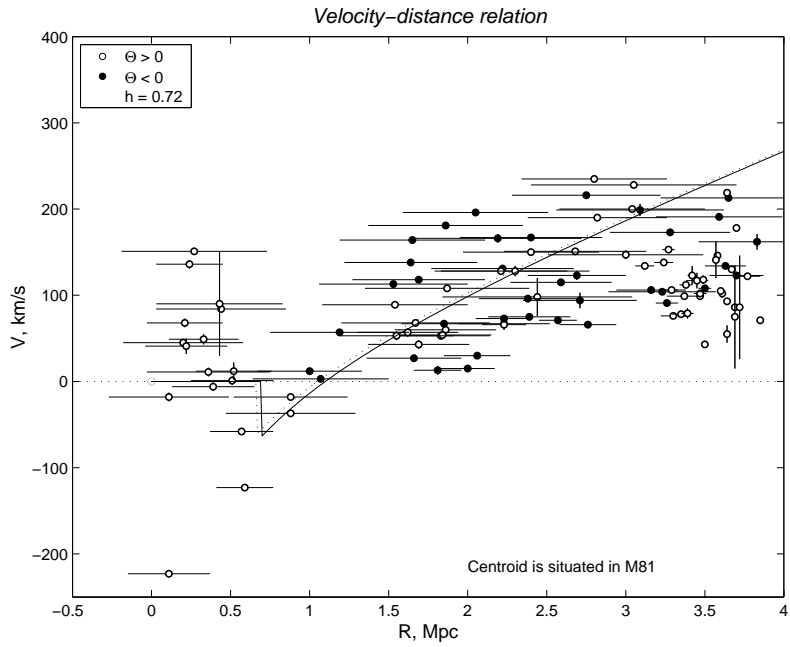}}
\vspace{15cm}
\caption{Hubble diagram for the  M~81 group with the center in  M~81}
\end{figure}

  \begin{figure}[h]
 \vbox{\includegraphics{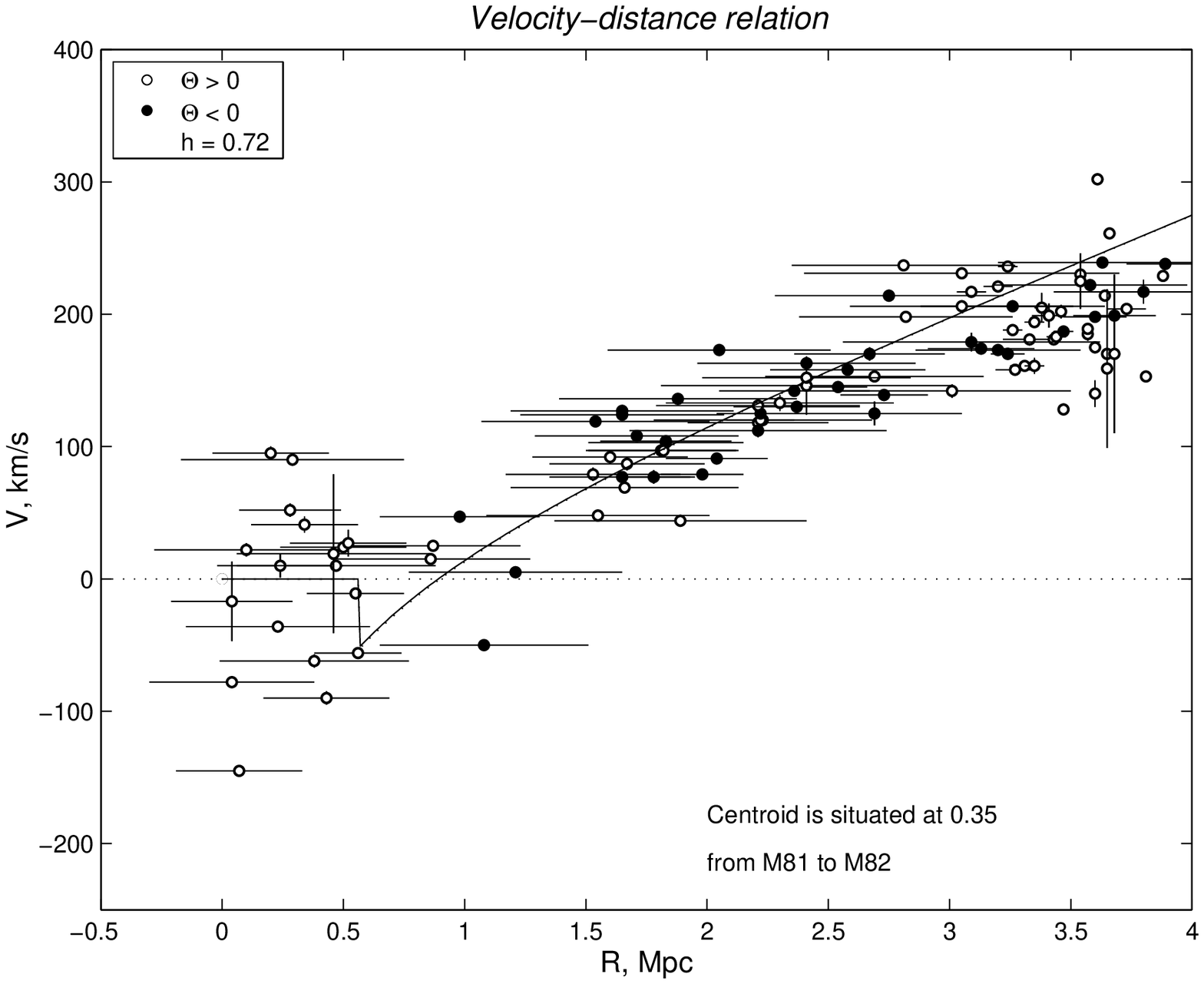}}
 \vbox{\includegraphics{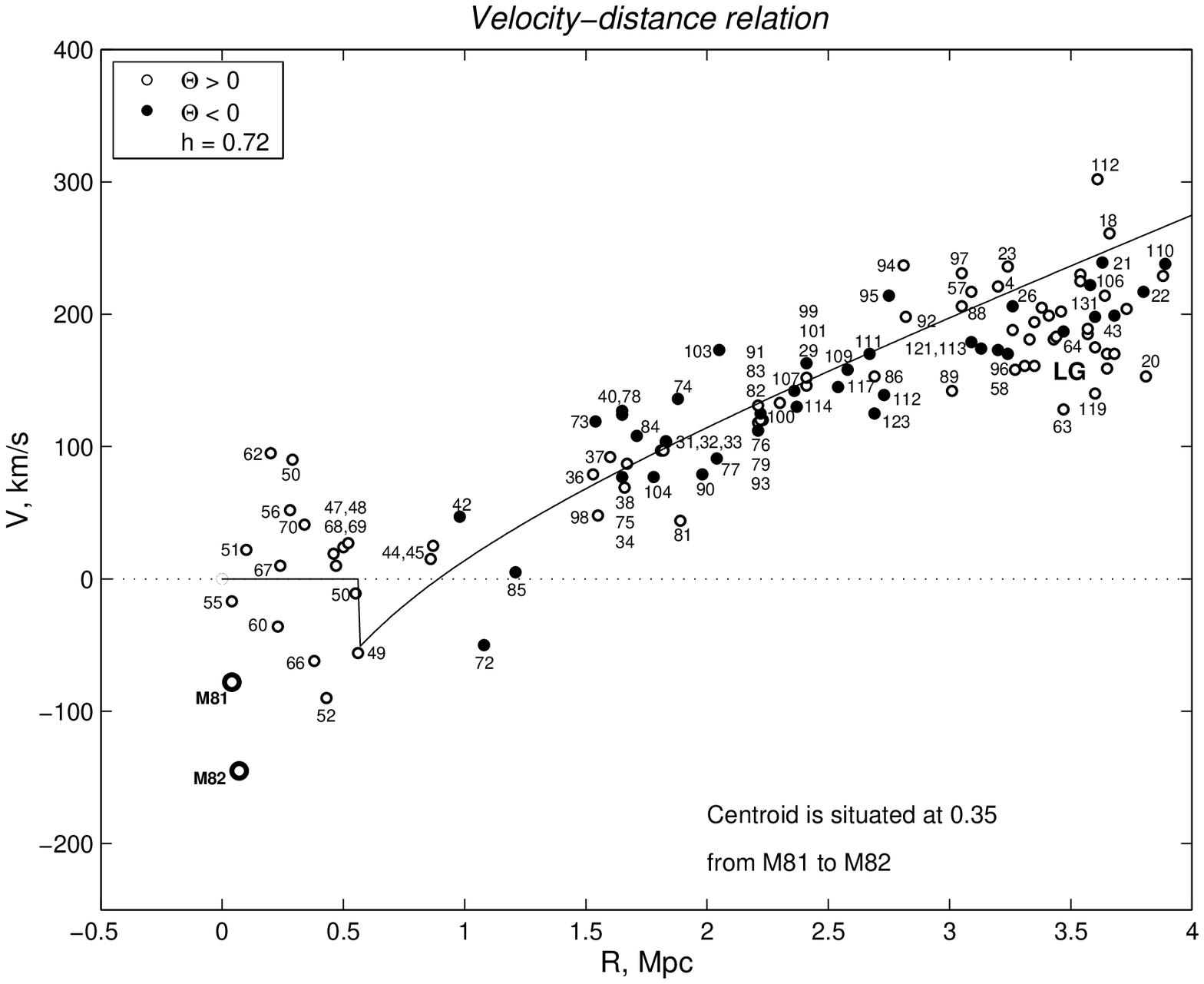}}
\vspace{20cm}
\caption{Hubble diagram for the M~81 group with the center of
masses at $x=0.35$ towards M~82.}
\end{figure}

\end{document}